\documentclass[fleqn,10pt]{wlscirep}
\usepackage[utf8]{inputenc}
\usepackage[T1]{fontenc}
\usepackage{amsmath, amssymb}
\usepackage{color, hyperref}
\usepackage[all]{hypcap}
\usepackage{graphicx}
\usepackage{xcolor}
\usepackage{placeins}
\usepackage[normalem]{ulem}
\usepackage{multicol}
\usepackage{float}
\usepackage{ifthen}
\usepackage{lineno}

\newlength{\figurewidth}
\newcommand{\Xe}[1]{Xe$^{#1+}$}

\newcommand{\trans}[2]{$#1 \rightarrow #2$}

\newcommand{\funit}{photons/\ensuremath{\mu}m\ensuremath{^2}}

\newcommand{\fluence}[2]{#1\ensuremath{\times}10\ensuremath{^{#2}}~\funit}

\title{Multiple-core-hole resonance spectroscopy with ultraintense X-ray pulses}

\author[1,2,*]{Aljoscha R{\"o}rig}
\author[3,$\dag$,*]{Sang-Kil Son}
\author[1]{Tommaso Mazza}
\author[1]{Philipp Schmidt}
\author[1]{Thomas M.\ Baumann}
\author[4]{Benjamin Erk}
\author[1,4,5]{Markus Ilchen}
\author[1]{Joakim Laksman}
\author[1,5]{Valerija Music}
\author[6]{Shashank Pathak}
\author[1]{Daniel E.\ Rivas}
\author[6]{Daniel Rolles}
\author[1]{Svitozar Serkez}
\author[1]{Sergey Usenko}
\author[2,3]{Robin Santra}
\author[1]{Michael Meyer}
\author[1,$\mathparagraph$]{Rebecca Boll}

\affil[1]{European XFEL, Holzkoppel 4, 22869 Schenefeld, Germany}
\affil[2]{Department of Physics, Universit{\"a}t Hamburg, Notkestr. 9-11, 22607 Hamburg, Germany} 
\affil[3]{Center for Free-Electron Laser Science CFEL, Deutsches Elektronen-Synchrotron DESY, Notkestr. 85, 22607 Hamburg, Germany}
\affil[4]{Deutsches Elektronen-Synchrotron DESY, Notkestr. 85, 22607 Hamburg, Germany}
\affil[5]{Institut f{\"u}r Physik und CINSaT, Universit{\"a}t Kassel, Heinrich-Plett-Str. 40, 34132 Kassel, Germany}
\affil[6]{J.\ R.\ Macdonald Laboratory, Department of Physics, Kansas State University, Manhattan, KS 66506, USA}

\affil[$\dag$]{sangkil.son@cfel.de}
\affil[$\mathparagraph$]{rebecca.boll@xfel.eu}
\affil[*]{these authors contributed equally to this work}

\begin{abstract}
Understanding the interaction of intense, femtosecond X-ray pulses with heavy atoms is crucial for gaining insights into the structure and dynamics of matter. One key aspect of nonlinear light-matter interaction was, so far, not studied systematically at free-electron lasers -- its dependence on the photon energy. Using resonant ion spectroscopy, we map out the transient electronic structures occurring during the complex charge-up pathways. Massively hollow atoms featuring up to six simultaneous core holes determine the spectra at specific photon energies and charge states. We also illustrate how the influence of different X-ray pulse parameters that are usually intertwined can be partially disentangled. The extraction of resonance spectra is facilitated by the fact that the ion yields become independent of the peak fluence beyond a saturation point. Our study lays the groundwork for novel spectroscopies of transient atomic species in exotic, multiple-core-hole states that have not been explored previously.
\end{abstract}

\begin{document}

\flushbottom

\maketitle

\thispagestyle{empty}

\section*{Introduction}

Extreme ultraviolet (XUV) and X-ray free-electron lasers (FELs) provide very intense pulses (10$^{12}$\,photons per pulse) with ultrashort pulse durations (a few tens of femtoseconds) that enable the absorption of more than one photon per atom or molecule~\cite{sorokin_photoelectric_2007, young_femtosecond_2010,rudenko_femtosecond_2017}. Such multiphoton interaction is a process of fundamental scientific interest because it enables studying the creation of (transient) ionic states of matter on timescales that were hitherto not accessible. In the X-ray regime, multiphoton inner-shell ionisation is dominated by sequential ionisation~\cite{jaeschke_interaction_2016}, while direct processes can become significant in the XUV regime~\cite{lambropoulos_multiple_2013,guichard_distinction_2014,Mazza_sensitivity_2015}. This extreme regime of nonlinear photon--matter interaction is important for various applications, such as single-particle imaging of biological macromolecules~\cite{neutze_potential_2000,chapman_X-ray_2019,bielecki_perspectives_2020,ekeberg_observation_2022}, Coulomb explosion imaging~\cite{boll_X-ray_2022, rudenko_femtosecond_2017,nagaya_ultrafast_2016}, the production of warm dense matter~\cite{vinko_creation_2012}, and the formation and control of plasmas in clusters or nanoparticles~\cite{kumagai_following_2018,gorkhover_femtosecond_2016,tachibana_nanoplasma_2015}. 

Far above inner-shell binding energies of a given atom and in the absence of saturation of the ionisation process, the yield of an ion charge state is proportional to $I^n$, with $I$ being the X-ray intensity ($=$\,number of photons per unit area and per unit time) and $n$ the average number of absorbed photons required to reach a given charge state~\cite{jaeschke_interaction_2016}. In contrast to multiphoton or tunnel ionisation using optical and infrared laser pulses~\cite{yamakawa_many-electron_2004}, the pulse duration was found to have a comparatively minor impact on the ion charge-state distributions in the X-ray regime~\cite{young_femtosecond_2010,jaeschke_interaction_2016}, except for charge states created predominantly via double-core-hole states~\cite{young_femtosecond_2010,hoener_ultraintense_2010, mazza_mapping_2020}. The X-ray fluence ($=$\,number of photons per unit area) was thus established as the most influential parameter for the resulting charge-state distributions. When the fluence becomes extremely high, the ion yields start to deviate from $I^n$, showing saturation. This saturation effect for ionisation is well known in the optical strong-field regime~\cite{walker_precision_1994,larochelle_non-sequential_1998,yamakawa_many-electron_2004}, and similar effects have previously been observed in FEL experiments in the XUV~\cite{sorokin_photoelectric_2007,makris_theory_2009} and X-ray~\cite{rudek_ultra-efficient_2012,rudek_resonance-enhanced_2013,rudek_relativistic_2018} regimes. However, to the best of our knowledge, the feature of saturation has not been exploited in high-intensity FEL applications yet. Here, we demonstrate how the deep saturation regime, in combination with a free tunability of the photon energy, facilitates a new type of ultra-high-intensity (transient) X-ray spectroscopy.

The photon-energy dependence of X-ray multiphoton absorption could, so far, only be investigated at very few selected photon energies due to the necessary re-tuning of the FEL at each photon energy. It was generally expected to map the decreasing photoabsorption cross section for increasing photon energy (far above any ionisation edges). For very high X-ray intensities, however, the opposite trend has been reported~\cite{rudek_relativistic_2018}, and transient resonances during charge-up~\cite{kanter_unveiling_2011} can lead to charge states significantly higher than expected at certain photon energies~\cite{rudek_ultra-efficient_2012,rudek_resonance-enhanced_2013,ho_theoretical_2014,ho_resonance-mediated_2015,toyota_interplay_2017,rudek_relativistic_2018}. Transient resonances have recently been found to influence molecular multiphoton ionisation~\cite{li_resonance-enhanced_2022} and to dramatically enhance scattering cross sections in X-ray diffraction imaging~\cite{kuschel_enhanced_2022}, but they can also cause increased radiation damage~\cite{ho_role_2020}. These observations illustrate the complex interplay between imaging and transient electronic structure, as well as the necessity for a careful choice of all X-ray pulse parameters. Isolating the influence of individual parameters on the experimental results can be difficult, but in the saturation fluence regime, one can disentangle them to a certain extent.

We present a joint theoretical and experimental study that sheds new light on the fundamental interaction of ultraintense soft-X-ray pulses with isolated atoms. The exceptionally high soft-X-ray pulse energies and the variable-gap undulators available at the Small Quantum Systems (SQS) instrument of the European X-ray Free-Electron Laser~\cite{decking_mhz-repetition-rate_2020} facilitate the extraction of resonance spectra for all charge states up to aluminium-like xenon (\Xe{41}) in the photon-energy range between 700 and 1700\,eV. In certain photon-energy regions, the rich structures in the resonance spectra result from the dominance of previously postulated massively hollow atoms~\cite{son_breakdown_2020} created by multiple inner-shell photoabsorptions. Specifically, atoms featuring up to six core holes are identified in the resonance spectra. While there are a few experimental examples of double-core-hole electron spectroscopy for atoms~\cite{mazza_mapping_2020} and molecules~\cite{santra_X-ray_2009, berrah_double-core-hole_2011}, such multiple-core-hole states have, to our knowledge, not been observed so far. 

\section*{Results}

\subsection*{Photon-energy-dependent charge-state distributions}
\label{sec:CSD}

\begin{figure}
\begin{center}
\includegraphics[width=\figurewidth]{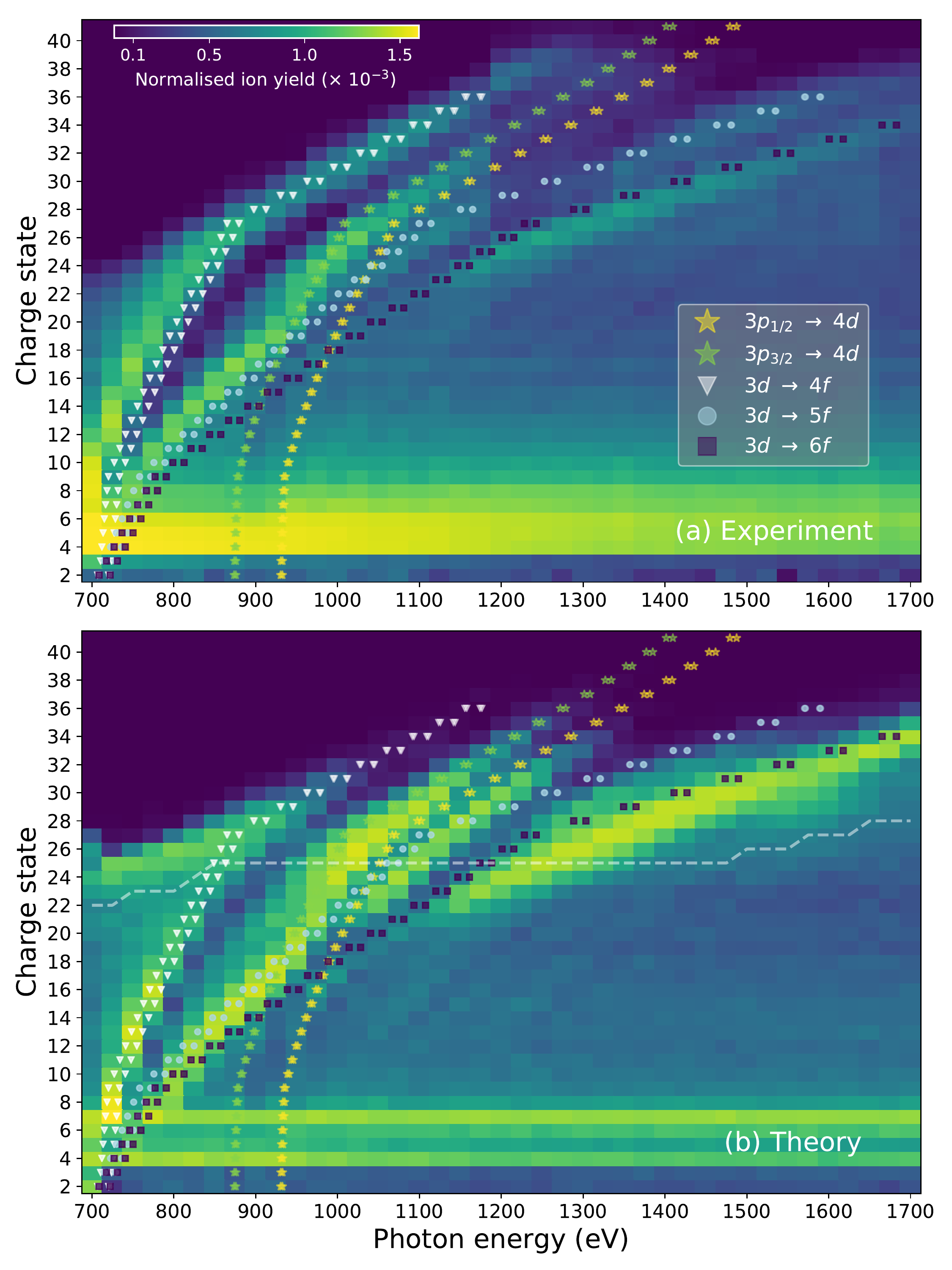}
\caption{Experimental (a) and calculated (b) xenon charge-state distributions as a function of photon energy. Coloured markers show selected resonant transition energies in the ground-state configuration of each charge state $q$\,--\,1. $(9.3\,\pm\,0.2)\,\times10^{12}$ photons per pulse on target were used in (a). The theoretical results in (b) were volume-integrated~\cite{toyota_xcalib_2019} with a peak fluence of \fluence{1.2}{12}. The white dashed line in (b) indicates the sequential direct one-photon ionisation limit (see text). The ion yields are normalised to the sum of all detected (calculated) ion yields for all photon energies in the experimental (theoretical) data set. The discontinuity at 1200\,eV in (a) is due to a change in the beamline optics (see Methods). The colour scale in (a) also applies to (b).}
\label{fig:exp_theo_map}
\end{center}
\end{figure} 

Figure~\ref{fig:exp_theo_map} shows measured (a) and calculated (b) xenon charge-state distributions~(CSDs) between 700 and 1700\,eV in steps of 25\,eV. This photon-energy range covers the ground-state orbital binding energies of the $3s$, $3p$, and $3d$ subshells of xenon. Charge states between \Xe{4} and \Xe{6} result from single-photon absorption~\cite{saito_multiple_1992}. Rich structures with several local maxima at significantly higher charge states are visible in both the experimental and calculated CSDs. They are generated by multiphoton absorption and shift towards higher charge states as the photon energy increases. Generation of the highest observed charge state, \Xe{41} at 1325\,eV, requires absorption of more than 30 photons (see Supplementary Fig.~\ref{fig:N_abs_ph}). 

During the X-ray pulse, xenon atoms charge up in a sequence of one-photon absorptions and relaxation events, particularly Auger-Meitner decay cascades. In the sequential direct one-photon ionisation limit~\cite{rudek_ultra-efficient_2012,toyota_interplay_2017,son_breakdown_2020} shown by the dashed white line in Fig.~\ref{fig:exp_theo_map}(b), the highest charge state $q$ is determined by the last ionic state $q\!-\!1$ that can be ionised with one photon from its electronic ground state. However, charge states significantly beyond this limit are observed in the CSDs. At specific photon energies and charge states, the increasing electron binding energies during the charge-up drive certain transitions between inner-shell orbitals and valence or Rydberg orbitals into resonance, thus leading to resonance-enabled or resonance-enhanced X-ray multiple ionisation~(REXMI)~\cite{rudek_ultra-efficient_2012, rudek_resonance-enhanced_2013}. 

The characteristic ion-yield maxima in Fig.~\ref{fig:exp_theo_map}, which shift as a function of photon energy, are a manifestation of the highly transient resonances. Relevant transition energies of the ground-state electronic configurations of charge state $q\!-\!1$ are indicated by coloured markers in both panels of Fig.~\ref{fig:exp_theo_map}. Changes in the slopes are related to the electronic configurations: The $O$ shell becomes empty at \Xe{8}, the $N$ shell at \Xe{26}, and the $3d$ subshell at \Xe{36}. In principle, accessible resonant excitations from the $3p$ and $3d$ subshells at given photon energies can be expected to lead to an increased yield of certain charge states. However, while the ion yield maxima in Fig.~\ref{fig:exp_theo_map} do indeed follow the general trend of the resonant transitions, it will become clear in the following that a correct assignment of the resonances requires to consider the complicated multiple ionisation dynamics during the FEL pulse in detail.

To this end, we performed \emph{ab initio} ionisation dynamics calculations using the \textsc{xatom} toolkit~\cite{son_impact_2011,jurek_xmdyn_2016}. We take into account volume integration in the calculations~\cite{toyota_xcalib_2019} (see Supplementary Discussion~\ref{app:volume}), because experimental ion spectra are always subject to a fluence distribution determined by the focused X-ray beam shape. As shown in Fig.~\ref{fig:exp_theo_map}(b), the theoretical CSDs reproduce the overall features of the experimental ion-yield maxima well. However, there are some noticeable differences. The experimental CSDs show a marked void between the two ion-yield branches around 800--1000\,eV, which is less pronounced in the theoretical CSDs, and a small shift of $\sim$25\,eV between experiment and theory, best seen between 700--800\,eV. We attribute this to the finite accuracy of the electronic structure method used in \textsc{xatom} (see Supplementary Table~\ref{table:E_shift} for details). Moreover, the highest charge states observed in the experiment around 1000--1400\,eV are missing in Fig.~\ref{fig:exp_theo_map}(b). As a consequence, the ion yield piles up for the intermediate charge states (\Xe{22} to \Xe{32}), thus enhancing structures related to the \trans{3d}{6f} excitation. We will see in the next subsection that the highest charge states can be reproduced by increasing the peak fluence in the calculation beyond the calibrated peak fluence. On the other hand, the theoretical CSDs show a structure around \Xe{22} to \Xe{26} at 700--800\,eV which is absent in the experimental data. This can be attributed to a reduced peak fluence at these photon energies in the experiment [see Fig.~\ref{fig:resonance_profiles}(c) and the corresponding text].

\subsection*{Resonance structures and peak-fluence dependence}
\label{sec:resonance_profile}

\begin{figure}
\begin{center}
\includegraphics[width=0.44\textwidth]{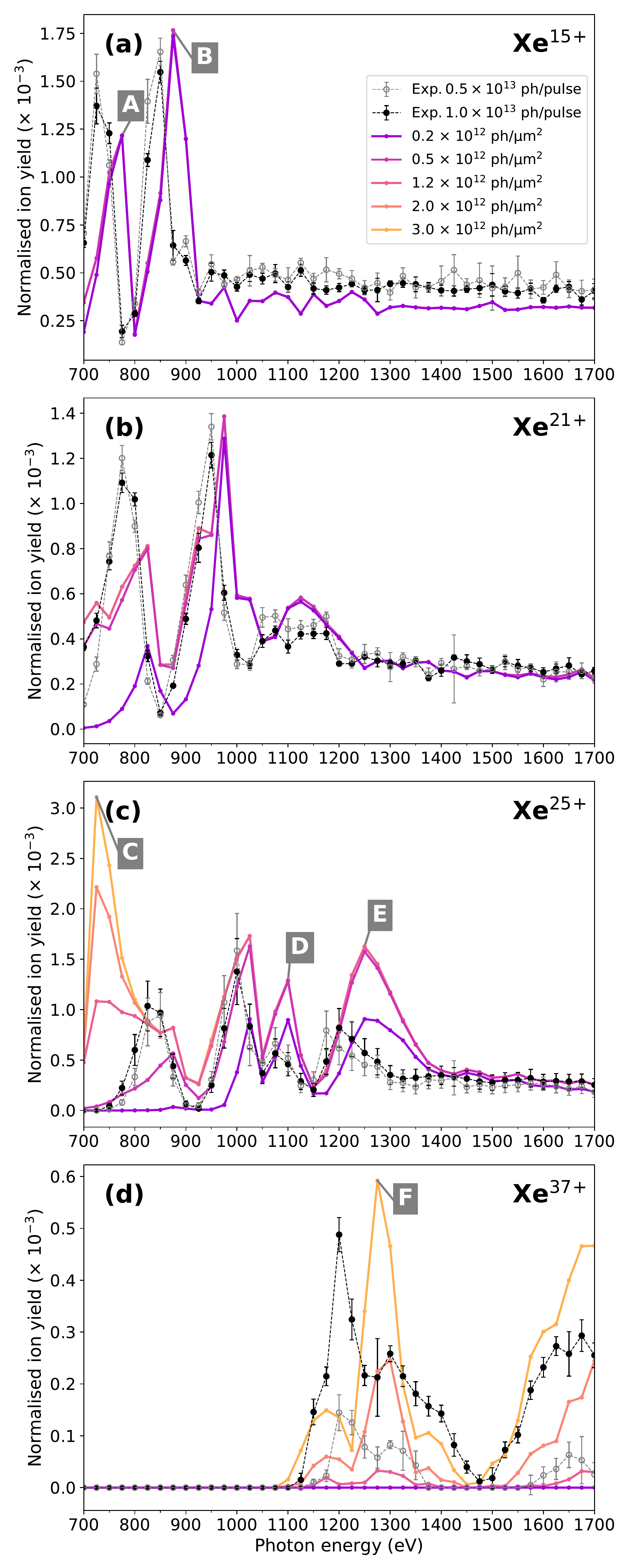}
\caption{Experimental and theoretical resonance spectra for four exemplary charge states of xenon. The experimental data for 1\,$\times$\,$10^{13}$ photons (black circles) in each panel correspond to horizontal lineouts of Fig.~\ref{fig:exp_theo_map}(a). A second data set recorded at 50\% X-ray transmission is also presented (grey open circles). Volume-integrated calculations are shown for different peak fluences in different colours. They are identical in some cases, such that only one or a few lines are visible. Several peaks are marked with A to F for further analysis in the main text and in Supplementary Discussion~\ref{app:transition}.}
\label{fig:resonance_profiles}
\end{center}
\end{figure}

Figure~\ref{fig:resonance_profiles} shows experimental (black and grey dashed lines) and theoretical (coloured lines) ion yields as a function of photon energy for four exemplary charge states, corresponding to lineouts of Fig.~\ref{fig:exp_theo_map}. 
These spectra display rich resonance structures which are absent in neutral xenon atoms~\cite{kanter_unveiling_2011} or are energetically inaccessible for low charge states. The most complex features are observed for the ``intermediate'' charge states, such as \Xe{25}~[Fig.~\ref{fig:resonance_profiles}(c)]. The overall trends, particularly the positions of minima and maxima, are well reproduced by theory.
A small shift, $\sim$25 eV, of the theoretical ion-yield maxima towards higher photon energies with respect to the experiment [best seen in Fig.~\ref{fig:resonance_profiles}(a)] is caused by inaccuracies in the transition energy calculation (see Supplementary Table~\ref{table:E_shift}).

Aside from the emergence of the resonance structures, one intriguing observation in Fig.~\ref{fig:resonance_profiles} is the influence of varying the peak fluence. Two experimental data sets at 50\% (grey) and 100\% (black) X-ray transmission are plotted. The peak fluence is generally assumed to be decisive for the resulting charge states in X-ray multiphoton ionisation. Indeed, for the high charge states, e.g., \Xe{37} [Fig.~\ref{fig:resonance_profiles}(d)], the ion yield increases nonlinearly for higher peak fluences. However, different behaviour is observed for lower charge states. The resonance spectrum for \Xe{15} [Fig.~\ref{fig:resonance_profiles}(a)] is completely insensitive to the peak fluence in both experiment and theory within the chosen fluence range. For \Xe{21} and \Xe{25} [Figs.~\ref{fig:resonance_profiles}(b) and (c)], the overall shapes of the resonance spectra remain unchanged for both experimental data sets, and the theoretical ion yields start to become sensitive to the peak fluence for photon energies lower than 975\,eV and 1400\,eV, respectively. In Fig.~\ref{fig:resonance_profiles}(c), a strongly fluence-dependent peak (labelled with `C') is visible around 725\,eV in the theory. Its absence in the experimental data is consistent with a peak fluence of approximately \fluence{0.3}{12} at this photon energy, matching our fluence calibration [see Supplementary Fig.~\ref{fig:num_photons}(b)].

\begin{figure}
\begin{center}
\includegraphics[width=0.9\figurewidth]{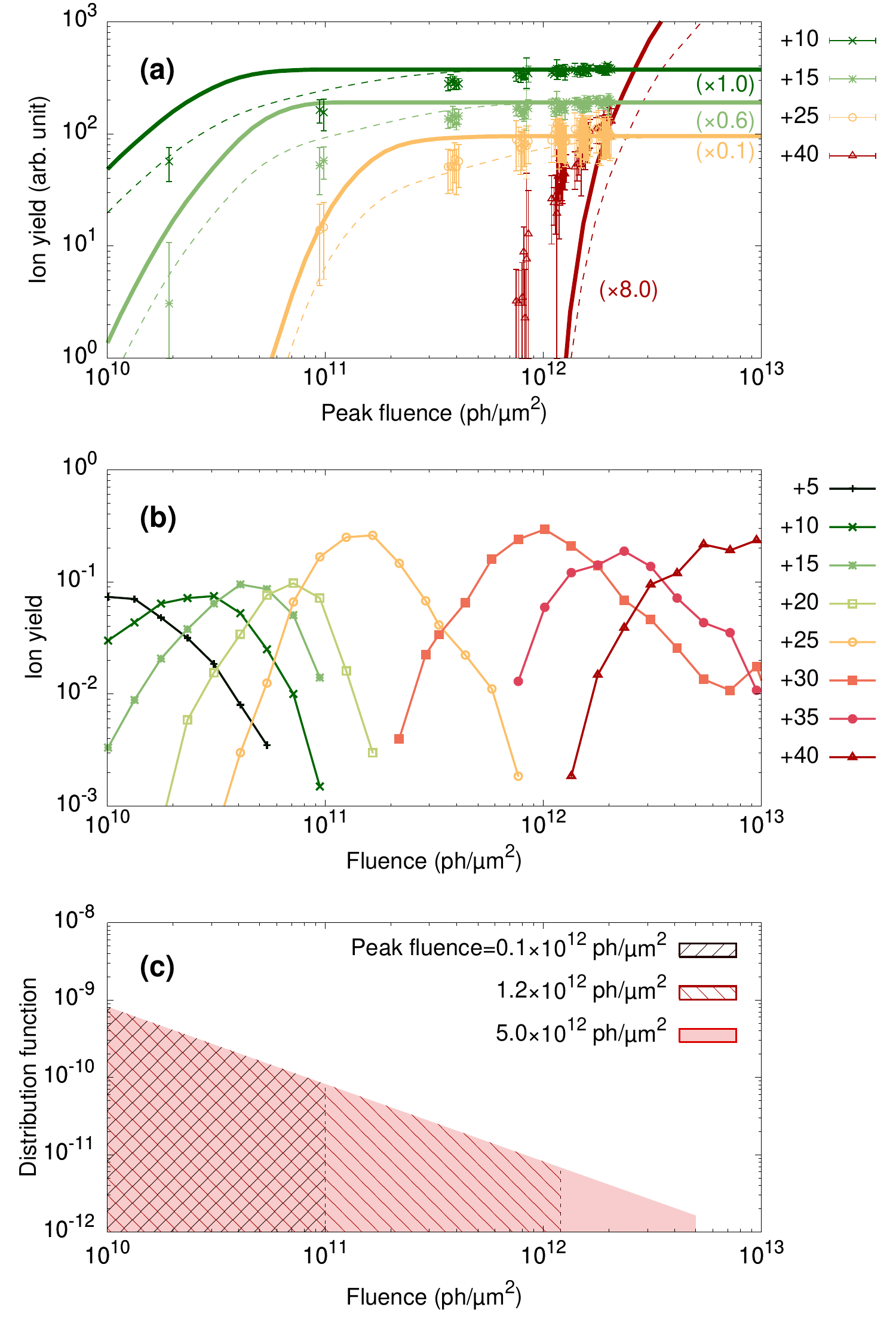}
\caption{(a) Ion yields of Xe at 1325\,eV as a function of peak fluence. Theoretical ion yields are volume-integrated with a single (solid lines) or double (dashed lines) Gaussian spatial profile (see Supplementary Discussion~\ref{app:volume} for details). Error bars of the experimental data include systematic and statistical errors (see Methods). 
The theoretical ion yields (both single and double Gaussian cases) are scaled to the experimental asymptotic values. The respective factors as specified next to the curves, with +10 being the reference.
(b) Computed Xe ion yields as a function of fluence without volume integration. (c) Fluence distribution function in the X-ray focus for different peak fluences, assuming a single Gaussian spatial profile.
}
\label{fig:ionyield_saturation}
\end{center}
\end{figure}

To further investigate the observed peak-fluence \mbox{(in-)dependence} of the resonance spectra, we plot the ion yields of several charge states of xenon at 1325\,eV photon energy as a function of peak fluence in Fig.~\ref{fig:ionyield_saturation}(a). Experimental and theoretical ion yields (after volume integration) increase nonlinearly for lower peak fluences but become flat beyond a certain saturation peak fluence. The onset of this saturation starts later for higher charge states and is not yet reached for \Xe{40}. We have multiplied the theoretical ion yields by individual factors (specified for each curve) to match the experimental data at the highest peak fluence (see also Supplementary Discussion~\ref{app:volume}). Without volume integration, as shown in Fig.~\ref{fig:ionyield_saturation}(b), we obtain the yield of a specific ion charge state as a function of fluence. This illustrates that every charge state is only generated in a relatively narrow fluence range [$f_1, f_2$], which corresponds to a narrow radial range [$r_2, r_1$] in a focused X-ray beam. For higher fluences, charge states are depleted because they become further ionised. Figure~\ref{fig:ionyield_saturation}(c) shows a fluence distribution function for different peak fluences (different numbers of photons per pulse), assuming a fixed single Gaussian spatial profile. It becomes clear that an increase of the peak fluence does not change the abundance of lower fluences. The area $[ = \pi(r_1^2-r_2^2)]$ covering the fluence range [$f_1, f_2$] in which a given charge state is generated remains constant when the number of photons per pulse increases~\cite{posthumus_dynamics_2004,toyota_xcalib_2019}, but the radii $r_1$, $r_2$ change depending on the peak fluence $F_0$: $r_i = \Delta \sqrt{ ( \ln \lbrace F_0 / f_i \rbrace ) / ( 4 \ln 2 ) }$, where $\Delta$ indicates the focal size (full width at half maximum, FWHM). In combination, Figs.~\ref{fig:ionyield_saturation}(b) and ~\ref{fig:ionyield_saturation}(c) explain why the volume-integrated ion yields in Fig.~\ref{fig:ionyield_saturation}(a) become constant. Exceptions are high charge states, for example, \Xe{40}, which are generated near the centre of the X-ray focus. The area in which they get created still increases when increasing the number of photons per pulse. Those charge states are not yet depleted in Fig.~\ref{fig:ionyield_saturation}(b), and saturation is not yet reached in Fig.~\ref{fig:ionyield_saturation}(a) for the maximum peak fluence available in the experiment. An onset of saturation has also been observed in previous multiphoton absorption studies in the soft~\cite{rudek_ultra-efficient_2012,rudek_resonance-enhanced_2013} and hard~\cite{rudek_relativistic_2018} X-ray regimes, but the peak fluence in those experiments was insufficient to reach a constant ion yield for many charge states. 

In summary, this demonstrates that the resonance spectra in Fig.~\ref{fig:resonance_profiles} are \emph{independent} of the X-ray fluence and the focal beam shape as soon as the saturation regime is reached. This makes multiphoton spectroscopy with ultraintense X-rays robust and insensitive, for example, to possible small variations of the X-ray focus size at different photon energies due to different beam divergence, and facilitates the unambiguous extraction of resonance features. The peak-fluence independence is a generic feature caused by the volume integration, and thus emerges for every sample and is independent of the target density used in the experiment. The results would be the same if only a single atom per pulse was located in the X-ray focal area.

\subsection*{Resonance assignment and multiple-core-hole analysis}
\label{sec:multiple-core-hole}
\begin{figure}
\begin{center}
\includegraphics[width=0.49\textwidth]{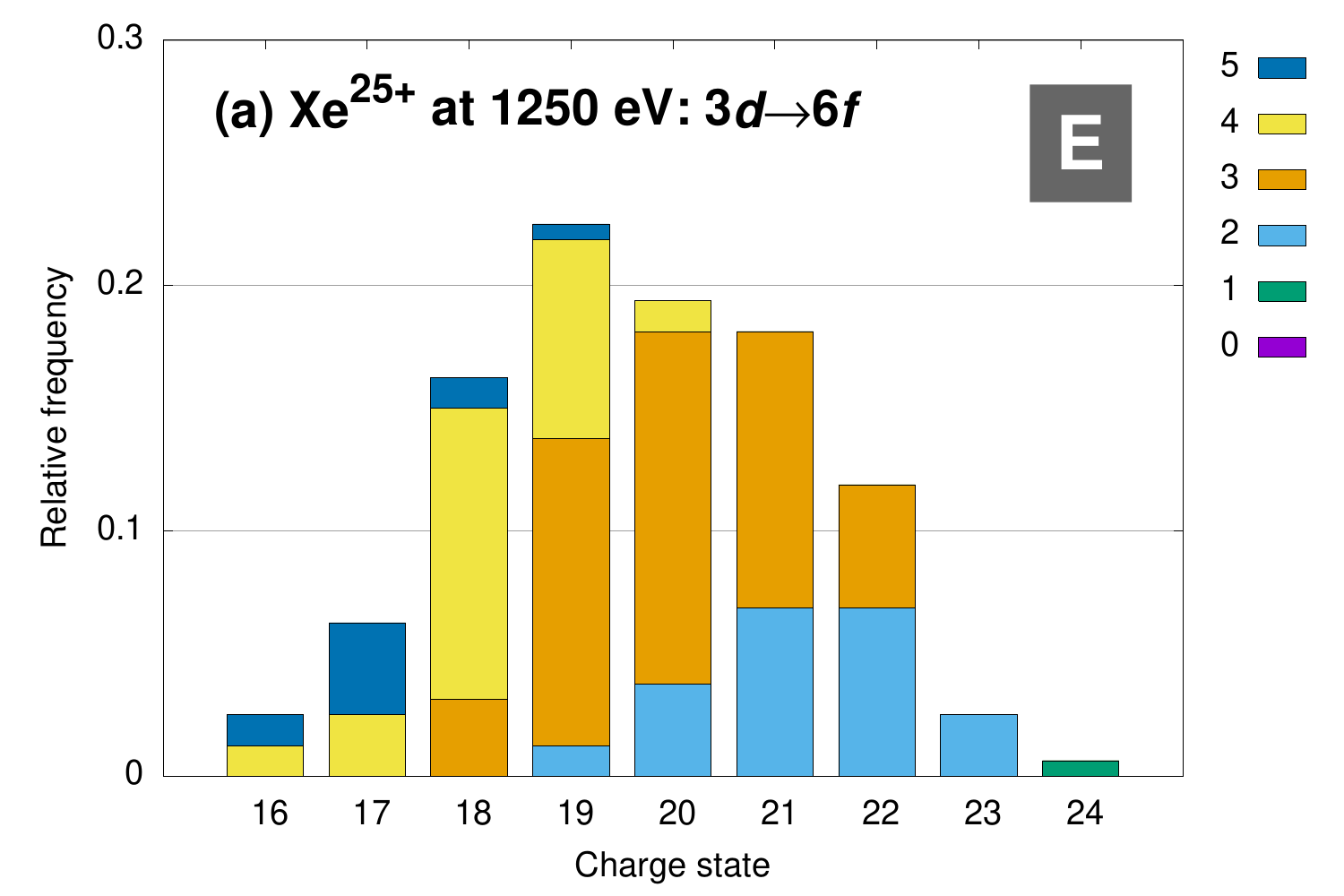}
\includegraphics[width=0.49\textwidth]{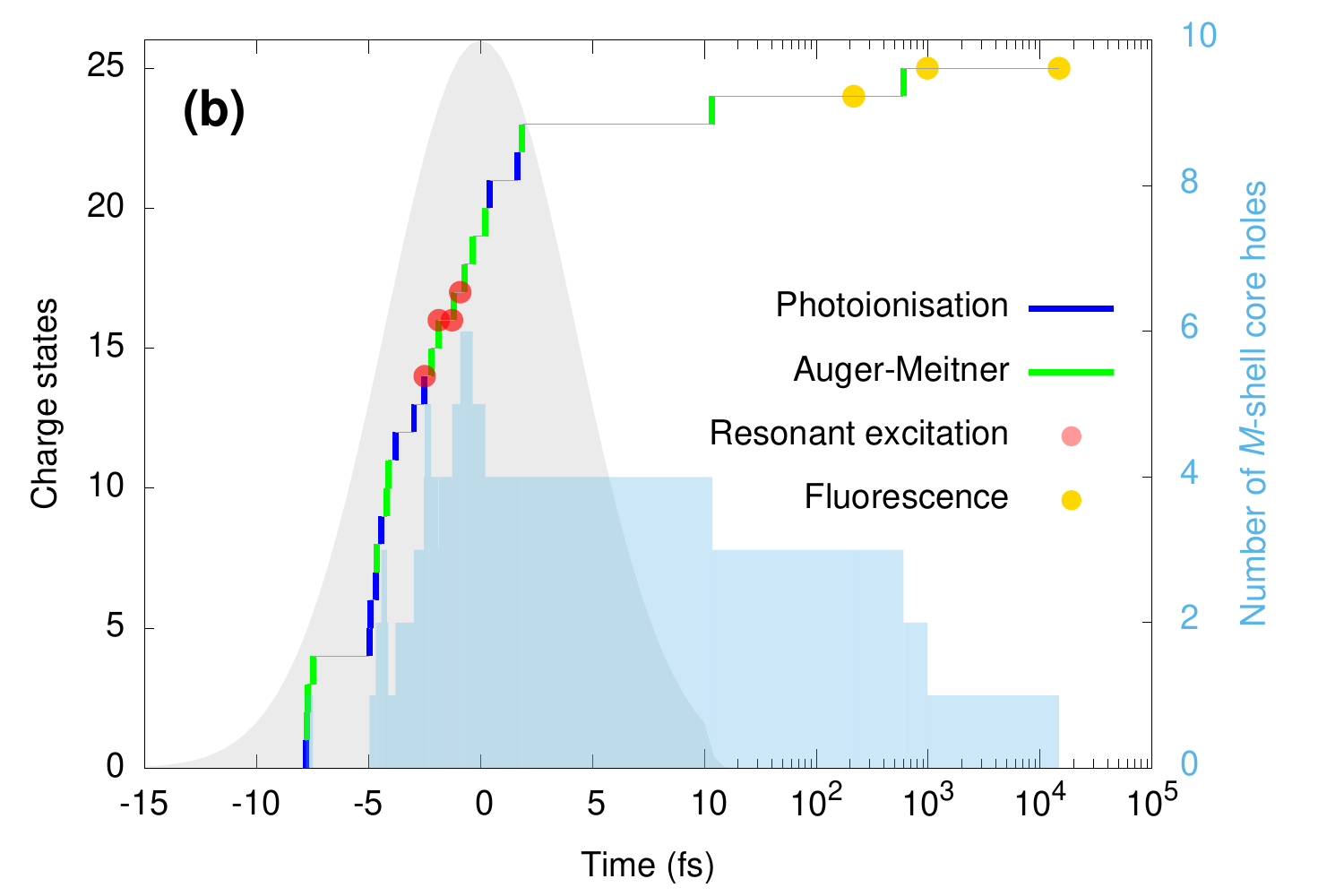}
\caption{Analysis of specific resonant excitations for \Xe{25} at 1250\,eV [peak E in Fig.~\ref{fig:resonance_profiles}(c)]. Peaks A--D, F are analysed in Supplementary Fig.~\ref{fig:histogram_last_resonances_all}. (a) A normalised histogram of the precursor charge states at which the last resonant excitation resulting in \Xe{25} occurs. The relative frequency shown is the number of Monte Carlo trajectories in which the \trans{3d}{6f} transition in a given charge state is the last resonant transition, divided by the total number of such Monte Carlo trajectories. Different colours indicate the relative number of $M$-shell core holes present at the time of the last resonant excitation. (b) An exemplary ionisation pathway corresponding to peak E, specifically involving five core holes in the precursor charge state of +17. For the given ionisation pathway, the time evolution of the number of $M$-shell core holes (right $y$ axis) is represented by the blue area. Note that the $x$ axis is on a logarithmic scale after 10\,fs.}
\label{fig:histogram_last_resonances}
\end{center}
\end{figure}

The peaks in the resonance spectra, such as those labelled A\,--\,F in Fig.~\ref{fig:resonance_profiles}, contain a plethora of information about the electronic structure and transient resonances. Based on the charge-state-dependent transition energies of certain ground-state resonances of charge state $q\!-\!1$ (coloured markers in Fig.~\ref{fig:exp_theo_map}), we can tentatively assign peaks A and B in \Xe{15} in Fig.~\ref{fig:resonance_profiles}(a) to the \trans{3d}{4f} and \trans{3d}{5f} transitions in \Xe{14}, respectively. However, this approach fails for the peaks in \Xe{25} in Fig.~\ref{fig:resonance_profiles}(c) -- the ion yield maxima do not match any ground-state $M$-shell transition energies of \Xe{24} (see Supplementary Table~\ref{table:peak_assignment}). The photon energy at which a given peak in the resonance spectra occurs not only depends on the charge state but also on the number of core holes and on the specific electron configuration, as illustrated in Supplementary Fig.~\ref{fig:cross_section}. In the following, we illustrate how the correct assignment for \Xe{25} can be carried out.

X-ray multiphoton ionisation dynamics are treated as Monte Carlo trajectories in our calculations~\cite{rudek_ultra-efficient_2012,fukuzawa_deep_2013}. In each trajectory, multiple resonant excitations can take place. The decisive transition for the structures in the resonance spectra of the final charge state is the last one. Therefore, we classified the calculated Monte Carlo trajectories through electronic configuration space according to the last resonant transition, and the peak is then identified as the most probable excitation.

In Fig.~\ref{fig:histogram_last_resonances}, we illustrate the trajectory analysis for peak~E as an example (see Supplementary Table~\ref{table:transition} and Supplementary Discussion~\ref{app:transition} for other peaks). At a fixed photon energy of 1250\,eV, Fig.~\ref{fig:histogram_last_resonances}(a) shows a normalised histogram of the precursor charge states at which the last resonant excitation resulting in a final charge state of \Xe{25} occurs. The histogram shows a wide distribution of precursor charge states, demonstrating that the last resonant excitation almost never happens at charge state \Xe{24}, but can take place at charge states as low as \Xe{16}. For this reason, an assignment of the resonance peaks is impossible without investigating the entire ionisation pathway with the help of ionisation dynamics calculations. After identifying the dominant transition, we find that the \trans{3d}{6f} transition causes the appearance of peak~E.

Furthermore, the analysis of trajectories allows us to retrieve transient core-hole states which are populated during the charge-up. In Fig.~\ref{fig:histogram_last_resonances}(a), colours indicate the number of core holes in the $M$ shell at the time of the last resonant excitation. Triple core holes are the dominant contribution, and up to quintuple core holes are detected. The last excitation creates one additional $M$-shell hole, i.e., sextuple core holes are present after the last excitation. Lower precursor charge states exhibit the highest number of core holes, and excitations without core holes do not exist in this case. This is an impressive demonstration of the complexity of the transient electronic structure emerging upon the interaction of matter with ultraintense X-ray pulses.

Figure~\ref{fig:histogram_last_resonances}(b) shows one exemplary ionisation pathway among the Monte Carlo trajectories corresponding to peak~E. It illustrates how 10 photoionisations, 4 resonant excitations, 15 Auger-Meitner decays, and 3 fluorescence decays lead to the final charge state \Xe{25}. The blue area indicates the time evolution of the number of $M$-shell core holes, reaching a maximum of 6. All individual electron configurations and their lifetimes involved in this specific ionisation pathway are listed in Supplementary Table~\ref{table:configuration}. Six core holes are formed at the peak of the X-ray pulse at \Xe{17} and \Xe{18}; their lifetimes are 0.75\,fs and 0.80\,fs, respectively. As the charge state increases, Auger-Meitner-type ionisation channels become unfavourable and their lifetime becomes longer because only a few valence electrons remain. In the Monte Carlo trajectory shown, the quadruple-core-hole state ($3d_{3/2}^{-2} 3d_{5/2}^{-2}$ from \Xe{20} to \Xe{23}) survives until 12\,fs and the triple-core-hole state ($3d_{3/2}^{-2} 3d_{5/2}^{-1}$ at \Xe{24}) lasts until 600\,fs. The last autoionisation takes place at 600\,fs and the remaining $M$-shell core holes are filled up via subsequent fluorescence events. The lifetimes of atomic species created via a sequence of core ionisation and excitation events vary by several orders of magnitude, depending on the charge state and the number of core holes. Thus, the multiple-core-hole resonance spectroscopy demonstrated in this work provides unique opportunities to examine exotic species with a wide range of lifetimes.

\section*{Discussion}

We have presented a novel kind of resonance spectroscopy using ultraintense femtosecond X-ray radiation. Stable operation of the free-electron laser over an energy range from 700 to 1700\,eV, in combination with advanced photon diagnostics and fast attenuation allowed us to maintain a constant number of photons per pulse on target during the scan. The resulting multiphoton resonance spectra unveil a wealth of structures, which can be assigned with the help of state-of-the-art theoretical calculations. We demonstrate that the resonance spectra become insensitive to the peak fluence above a certain saturation peak fluence, which allows isolating the effects of X-ray beam parameters that are otherwise intertwined with the dominant fluence dependence. In our case, this facilitates the characterisation of the transient resonant excitations during the charge-up.  

Transient multiple-core-hole states are found to be crucial for explaining the peaks in the resonance spectra, and some peaks are even exclusively formed via two or more core holes. This opens up new avenues for multiple-core-hole (resonance) spectroscopy with ultraintense and ultrashort XFEL pulses. We demonstrate that extremely short-lived, as well as unusually long-lived, highly charged ions in exotic electronic configurations can be created and probed simultaneously through interaction with intense soft-X-ray pulses. Such unusual atomic species may also be formed via collisions in outer space~\cite{cumbee_charge_2017,gu_detection_2022}, which could be potential candidates for unidentified X-ray emission lines in astrophysics~\cite{hell_highly_2020,ezoe_high-resolution_2021}. 

In principle, electron spectroscopy can provide additional information on the transient multiple-core-hole states. However, in practice, their interpretation would be hampered by many overlapping emission lines. Furthermore, at such high degrees of ionisation (up to 41 electrons emitted from a single atom), Coulomb interaction among the ejected electrons would inevitably broaden the emission lines. Ion spectroscopy is therefore advantageous for investigating multiphoton multiple ionisation dynamics at ultraintense X-ray fluences. High-resolution fluorescence measurements~\cite{agaker_1-d_2023}, while only representing a minority of all involved electronic transitions, will provide valuable complementary information to further benchmark theory in the future. Novel seeding~\cite{hemsing_enhanced_2020,amann_demonstration_2012} and two-colour schemes~\cite{serkez_opportunities_2020} may be exploited to systematically study the evolution of transient resonances both in the spectral and temporal domains, and upcoming possibilities of tuning the pulse duration in addition to the photon energy will provide exciting options for other new spectroscopy techniques.

\section*{Methods}


\subsection*{Experiment}
\label{sec:exp}

The experiment was carried out at the SQS scientific instrument at the European XFEL~\cite{decking_mhz-repetition-rate_2020}. Isolated xenon atoms were irradiated with ultraintense, femtosecond soft-X-ray pulses. Xenon gas was introduced into the Atomic-like Quantum Systems (AQS) experimental station through an effusive needle at an ambient pressure of around 3\,$\times10^{-8}$\,mbar. The ions resulting from the interaction with the X-ray pulses were recorded by a high-resolution ion time-of-flight (TOF) spectrometer~\cite{de_fanis_high-resolution_2022} and fast analogue-to-digital converters with a resolution of 0.25\,ns. 

The photon energy was scanned over a range of 1\,keV (700\,--\,1700\,eV) in 25\,eV steps while maintaining a constant high number of photons per pulse (0.5 or 1\,$\times10^{13}$) throughout the scan. This was facilitated by i) the tunable variable-gap SASE3 undulators~\cite{tschentscher_photon_2017}, ii) the high energies per pulse (2.3\,--\,6.4\,mJ), and iii) a fast-responding, 15-meter-long gas attenuator filled with nitrogen gas~\cite{dommach_photon_2021}. The attenuator was adjusted for every photon energy to compensate for the change in initial pulse energy and used to perform fluence scans over more than two orders of magnitude at fixed photon energies. An additional gas monitor detector~\cite{tiedtke_gas_2008} was installed downstream of the interaction chamber for this experiment, to characterise the number of photons on target in parallel, as described in Ref.~\citeonline{baumann_harmonic_2023}. Peak fluences $>$\,\fluence{1}{12} were achieved due to the few-micron focus created by the SQS focusing optics~\cite{mazza_beam_2023} (see also Supplementary Fig.~\ref{fig:num_photons}). 

The calculated upper limit of the X-ray pulse duration was 25 fs based on the electron bunch charge of 250\,pC in the accelerator. To date, no direct measurement of the X-ray pulse duration has been carried out at the European XFEL, but indirect measurements suggest that the pulse duration can be approximately 10\,fs\cite{khubbutdinov_high_2021}. The European XFEL operated at a 10\,Hz repetition rate, at which it provided bursts of electron pulses with an inter-pulse frequency of 2.25\,MHz. We used every 32nd of these pulses to produce photons. The X-ray pulses thus had a spacing of 14.2\,$\mu$s, chosen to avoid overlapping ion spectra. Overall, we received 250\,--\,350 X-ray pulses per second. The bandwidth was measured with a grating spectrometer in advance of the experiment and was 1\,--\,2\% for all photon energies.

\subsection*{Data analysis}

In order to obtain quantitative charge-state distributions (CSDs), several analysis steps were carried out. Single-shot ion traces were obtained by cutting the recorded traces in 14.2\,$\mu$s segments -- the time interval between FEL pulses in the burst. Subsequently, we applied a filter to only analyse FEL shots with pulse energies within one standard deviation of the average pulse-energy distribution of the pulse train. For each analysis step, a systematic error estimate is carried out, as detailed in the following. The statistical error is extracted by splitting the data set at each photon energy and peak fluence bin into four subsets and evaluating the deviation between the CSDs of the four subsets.

\textbf{Isotope deconvolution:} 
Up to \Xe{15}, all seven stable isotopes of xenon were resolved without superimposing isotopes of higher charge states in the TOF spectrum. For charge states higher than \Xe{15}, some peaks of higher mass and higher charge, $m_i/q_i$, start overlapping with those of lower mass and lower charge, $m_{i-1}/q_{i-1}$. We, therefore, applied a deconvolution algorithm to extract the ion yields of all charge states, using the known isotope distribution as an input. We use the convolution theorem by inverse Fourier transformation of the Fourier-transformed TOF signal (S) divided by the Fourier-transformed isotope structure (R) extracted from the averaged TOF signal. S and R are on a logarithmic scale before the Fourier transformation is applied, making the isotope peaks equidistant for all charge states within the isotope structure. Otherwise, the distances would scale with the inverse of the charge state $q$ and the deconvolution algorithm could not be applied. The algorithm's error is estimated using the difference between the input and output of the deconvolution procedure on a simulated xenon TOF spectrum.

\textbf{Charge-state-dependent detector efficiency:} The per-shot ion counts ($\approx$90) were too high to safely neglect double counts in the same mass-over-charge peak. Thus, we did not attempt to convert the raw data to individual ion counts, e.g., via software constant fraction discrimination, but instead applied a charge-state-dependent correction factor to the ion yield of each charge state, accounting for the increase in average microchannel-plate (MCP) signal height for higher charge states~\cite{lienard_performance_2005, mroz_micro_1999}. This correction factor~\cite{gilmore_ion_2000} was obtained by recording a separate calibration data set at significantly reduced gas pressure, for which constant fraction discrimination could be applied, and extracting the peak pulse height of the pulse-height distribution for each charge state.
The error of the correction is estimated by the difference between the counted spectra and the corrected spectra of the calibration data set.

\textbf{Background subtraction:} Background signal from ionisation of residual gas mainly consists of oxygen ions from water. By evaluating the pulse-height distribution of xenon isotope 132 in counting mode, the oxygen contribution can be estimated through its distinctly lower MCP pulse heights in comparison to highly charged xenon ions with the same flight time. This fraction is subtracted from the ion yields of \Xe{25}, \Xe{33} and \Xe{41}, which overlap with O$^{3+}$, O$^{4+}$, and O$^{5+}$, respectively. The error estimation is based on the difference between spectra with and without background subtraction.

\textbf{Target density normalisation:} The two data sets above and below 1200\,eV photon energy, as well as the data at 50\% transmission, were recorded with slightly different gas pressures. Therefore, the ion yields were normalised to the gas pressure measured by an ion gauge in the interaction chamber parallel to the data recording.

\subsection*{Modelling}
\label{sec:theo}

To interpret the experimental data and to elucidate the underlying ionisation mechanisms, \emph{ab initio} ionisation dynamics calculations using the \textsc{xatom} toolkit~\cite{son_impact_2011,jurek_xmdyn_2016} were performed. \textsc{xatom} has recently been extended to incorporate resonance and relativistic effects~\cite{toyota_interplay_2017, rudek_relativistic_2018}. For any given electronic configuration of Xe ions, the electronic structure was calculated on the basis of the Hartree-Fock-Slater method, implementing first-order relativistic energy corrections. The atomic data, including photoabsorption cross sections, Auger-Meitner (including Coster-Kronig) rates, and fluorescence rates, were calculated in leading-order perturbation theory.

Using a rate-equation approach~\cite{rohringer_x-ray_2007,young_femtosecond_2010}, the X-ray multiphoton ionisation dynamics~\cite{jaeschke_interaction_2016} were simulated by solving a set of coupled rate equations with calculated atomic data. For the given range of photon energies, the ionisation dynamics of Xe are mainly initiated by $M$-shell ($n=3$) ionisation. The number of coupled rate equations, which is equivalent to the number of electronic configurations that are formed by removing zero, one, or more electrons from initially occupied subshells ($n \geq 3$) of Xe ions and placing them into $(n,l)$-subshells ($n \leq 30$ and $l \leq 7$), is $\sim 4.2\times10^{60}$ (see Refs.~\citeonline{son_breakdown_2020,ho_resonance-mediated_2015}). To handle such an enormous number of rate equations, we used a Monte Carlo on-the-fly approach~\cite{fukuzawa_deep_2013}. We assumed no contribution from direct (nonsequential) two-photon absorption~\cite{doumy_nonlinear_2011} and no effect due to the chaoticity of FEL self-amplified spontaneous emission (SASE) pulses~\cite{rohringer_x-ray_2007}. Higher-order many-body processes such as double photoionisation via shakeoff or knockout mechanisms~\cite{schneider_separation_2002} and double Auger-Meitner decay~\cite{kolorenc_collective_2016} were not included.

We used an energy bandwidth of 1\% and a Gaussian temporal profile with a 10-fs FWHM in all our calculations. Unless otherwise noted, the theoretical results were volume-integrated~\cite{toyota_xcalib_2019} with a peak fluence of \fluence{1.2}{12}, which was obtained as the mean value of the calibrated peak fluences for 1200--1700\,eV (see Supplementary Fig.~\ref{fig:num_photons}), and a single Gaussian spatial profile was used in the volume integration. All X-ray beam parameters were kept constant while the photon energy was varied.

\section*{Data availability}
Data recorded for the experiment at the European XFEL are available at \href{https://doi.org/10.22003/XFEL.EU-DATA-002310-00}{https://doi.org/10.22003/XFEL.EU-DATA-002310-00}.

\bibliography{2023_Roerig_Xenon}


\section*{Acknowledgements}

We acknowledge European XFEL in Schenefeld, Germany, for the provision of X-ray free-electron laser beam time at the SQS instrument and would like to thank the staff for their assistance. We also thank the operators and the run coordinators at DESY for their commitment and patience in tuning and performing the wide photon energy scans for the first time. 
We thank Jos\'e R.\ Crespo L\'opez-Urrutia and Rui Jin for their help with \textsc{fac} calculations. 
We thank Kai Tiedtke, Andrey Sorokin, Fini Jastrow and Yilmaz Bican for providing the gas monitor detector downstream of the instrument. We also thank Theophilos Maltezopoulos for his support. 
A.R.\ and M.M.\ acknowledge funding by the Deutsche Forschungsgemeinschaft (DFG, German Research Foundation)--SFB-925--project 170620586.
M.I., V.M.\ and Ph.S.\ acknowledge funding from the Volkswagen foundation for a Peter Paul Ewald-fellowship.
S.P.\ and D.R.\ were supported by the US Department of Energy, Office of Science, Office of Basic Energy Sciences, under contract no.~DE-FG02-86ER13491. 

\section*{Author contributions statement}

R.B.\ and S.-K.S.\ conceived the beam time. A.R., S.-K.S., T.M., P.S., T.M.B., B.E., M.I., J.L., V.M., S.P., D.E.R., D.R., S.S., S.U., M.M., and R.B. carried out the experiment. 
A.R., with help of P.S., T.M., and R.B., analysed the data. 
S.-K.S.\ carried out the calculations using the \textsc{xatom} toolkit (developed by S.-K.S.\ and R.S.). 
A.R., R.B., S.-K.S., R.S., and M.M. interpreted the results and wrote the manuscript with input from all authors.

\section*{Additional information} 

\subsection*{Correspondence} Requests should be addressed to Rebecca Boll and Sang-Kil Son.

\subsection*{Competing interests} 
The authors declare no competing interests.


\newpage
\renewcommand{\thesubsection}{S\arabic{subsection}}
\renewcommand{\thefigure}{S\arabic{figure}}
\renewcommand{\thetable}{S\arabic{table}}
\setcounter{subsection}{0}
\setcounter{table}{0}
\setcounter{figure}{0}
\begin{appendix}
\section*{Supplementary Information}
Aljoscha R\"orig \emph{et al.}, Multiple-core-hole resonance spectroscopy with ultraintense X-ray pulses
\section*{Supplementary Discussion}

\subsection{Fluence determination and volume integration}
\label{app:volume}

The experimental data are subject to a fluence distribution due to the focus of the X-ray pulse. This leads to the so-called focal volume effect~\cite{posthumus_dynamics_2004} or volume integration~\cite{toyota_xcalib_2019}---the fact that, in addition to the maximum (peak) fluence, regions of lower fluences also contribute to the measured ion signal. For a quantitative comparison between theory and experiment, it is important to take into account volume integration when computing theoretical data using the experimental fluence distribution in the interaction volume. We employ an established calibration procedure~\cite{toyota_xcalib_2019} using ion yields of argon recorded under identical experimental conditions. 

By using an extended version of \textsc{xcalib}~\cite{breckwoldt_machine-learning_2023}, which can employ a series of pulse-energy data points, we extracted a focal spot size as a function of photon energy as shown in Supplementary Fig.~\ref{fig:num_photons}(b). A single Gaussian spatial profile was used to model the focal shape in the two dimensions perpendicular to the beam propagation. The fluence distribution along the beam propagation direction was assumed to be constant, because the acceptance length of the spectrometer along the FEL propagation axis, approximately $\pm 1.5$\,mm, was shorter than the measured Rayleigh length, $\sim$3--6\,mm.

The experimental data shown in Fig.~\ref{fig:exp_theo_map}(a) were recorded in two separate sets for which two different beamline configurations were used: for 700\,--\,1175\,eV, the ``low-energy premirror'' (LE) with a 13\,mrad offset mirror chicane was used, whereas data for photon energies of 1200\,--\,1700\,eV were recorded with the ``high-energy premirror'' (HE) and a 9\,mrad offset mirror chicane (see Ref.~\citeonline{mazza_beam_2023} for details). The pulse energy was measured on a shot-to-shot basis by two gas monitor detectors after the undulators and downstream of the experiment, respectively. The beamline transmission is well characterised~\cite{mazza_beam_2023}. The number of photons on target was kept constant throughout the entire scan ($\sim 9.5\times10^{12}$\,photons), as shown in Supplementary Fig.~\ref{fig:num_photons}(a), by changing the transmission of the SASE3 gas absorber and monitoring the downstream gas-monitor detector. However, the change of beamline configuration resulted in a lower peak fluence for the LE data set, because the focus size was slightly larger. 

The theoretical data shown in Fig.~\ref{fig:exp_theo_map}(b) were volume-integrated with a peak fluence of \fluence{1.2}{12}, which corresponds to the mean of the calibrated peak fluences for 1200\,--\,1700\,eV. We chose the HE data set because Ar ionisation dynamics at low photon energies (700\,--\,900\,eV) are influenced by resonances and thus the Ar calibration becomes susceptible to other X-ray parameters, e.g., the spectral bandwidth. Assuming a single Gaussian profile, the calibrated focal size is $\Delta$=2.7\,$\mu$m (FWHM). The measured number of photons per pulse, $N_\text{ph}$, is converted into the peak fluence, $F_0$, as: $F_0 = (4\ln 2 / \pi) \cdot N_\text{ph} / \Delta^2$. 

In Fig.~\ref{fig:ionyield_saturation}(a), we have multiplied the theoretical ion yields by individual multiplication factors to match the experimental data at the highest fluence. While the overall shapes of the resonance spectra in Fig.~\ref{fig:resonance_profiles} are in good agreement, we see in Fig.~\ref{fig:ionyield_saturation}(a) that the absolute yields of intermediate charge states such as \Xe{25} are overestimated, while high charge states such as \Xe{40} are underestimated.
In order to test whether these inconsistent scaling factors could be caused by improper volume integration, we compare two theoretical data sets with single (solid lines) and double (dashed lines) Gaussian spatial profiles used in the volume integration. The latter is a typical way to accommodate a low-fluence background tail in the focused beam~\cite{mazza_beam_2023,toyota_xcalib_2019}. At a photon energy of 1325\,eV, the calibration procedure using pulse-energy-dependent Ar CSDs for the double Gaussian profile provides a fluence ratio of $f_r$=0.55 and a width ratio of $w_r$=2.0, and a focal size of the first Gaussian is $\Delta_1$=1.72\,$\mu$m (FWHM). While the agreement between theory and experiment in the low peak-fluence regime is better (except for \Xe{40}), the volume integration with the double Gaussian shape does not resolve the incongruous scaling factors for individual ion yields (the same factors are used for the single and double Gaussian cases). This suggests that a careful validation of the atomic data employed in our calculation is required, especially for multiply excited ions in the soft-X-ray regime. Further developments can be proposed, for example, the inclusion of higher-order many-body processes~\cite{schneider_separation_2002,kolorenc_collective_2016}, the chaoticity of SASE pulses~\cite{rohringer_x-ray_2007}, and coherence effects~\cite{kanter_unveiling_2011,rohringer_strongly_2012,li_coherence_2016}. The resonance positions are expected to profit from improved electronic structure theory~\cite{budewig_theoretical_2022}, which affects resonant ionisation dynamics~\cite{budewig_state-resolved_2023}. We note that a possible low-fluence pedestal in the experiment would not affect the observed peak-fluence insensitivity of the resonance spectra, because the double Gaussian curves in Fig.~\ref{fig:ionyield_saturation}(a) also show saturation.

\subsection{Analysis of resonant transitions}
\label{app:transition}

Supplementary Table~\ref{table:peak_assignment} shows peak assignments based on the ground-state transition energies of charge state $q\!-\!1$.
$E_\text{peak}$ corresponds to the ion yield maxima of the theoretical data shown in Fig.~\ref{fig:resonance_profiles}.
For each charge state $q$, the transition energies of six different resonant excitations are listed, which are obtained from the ground-state calculation for charge state $q\!-\!1$.
They are located in the same row as the closest $E_\text{peak}$.
Note that there is no transition from $3d$ at \Xe{37}, because $3d$ is empty for $q\!>\!+36$.
In this way, some of the peaks can be assigned: for example, 775\,eV and 875\,eV at \Xe{15} correspond to the transitions \trans{3d}{4f} and \trans{3d}{5f} of \Xe{14}, respectively, and 1250\,eV at \Xe{25} to \trans{3d}{7f} of \Xe{24}.
However, this ground-state-based assignment fails for many other peaks: for example, 1100\,eV at \Xe{25} is far from any ground-state transition energies of \Xe{24}.
The nearest transition is \trans{3d}{5f} (1060\,eV) and the next one is \trans{3p_{1/2}}{4d} (1050\,eV), both of which are separated from $E_\text{peak}\!=\!1100$\,eV by $\geq$40\,eV.

\begin{figure}[H]
\centering
\includegraphics[width=0.8\figurewidth]{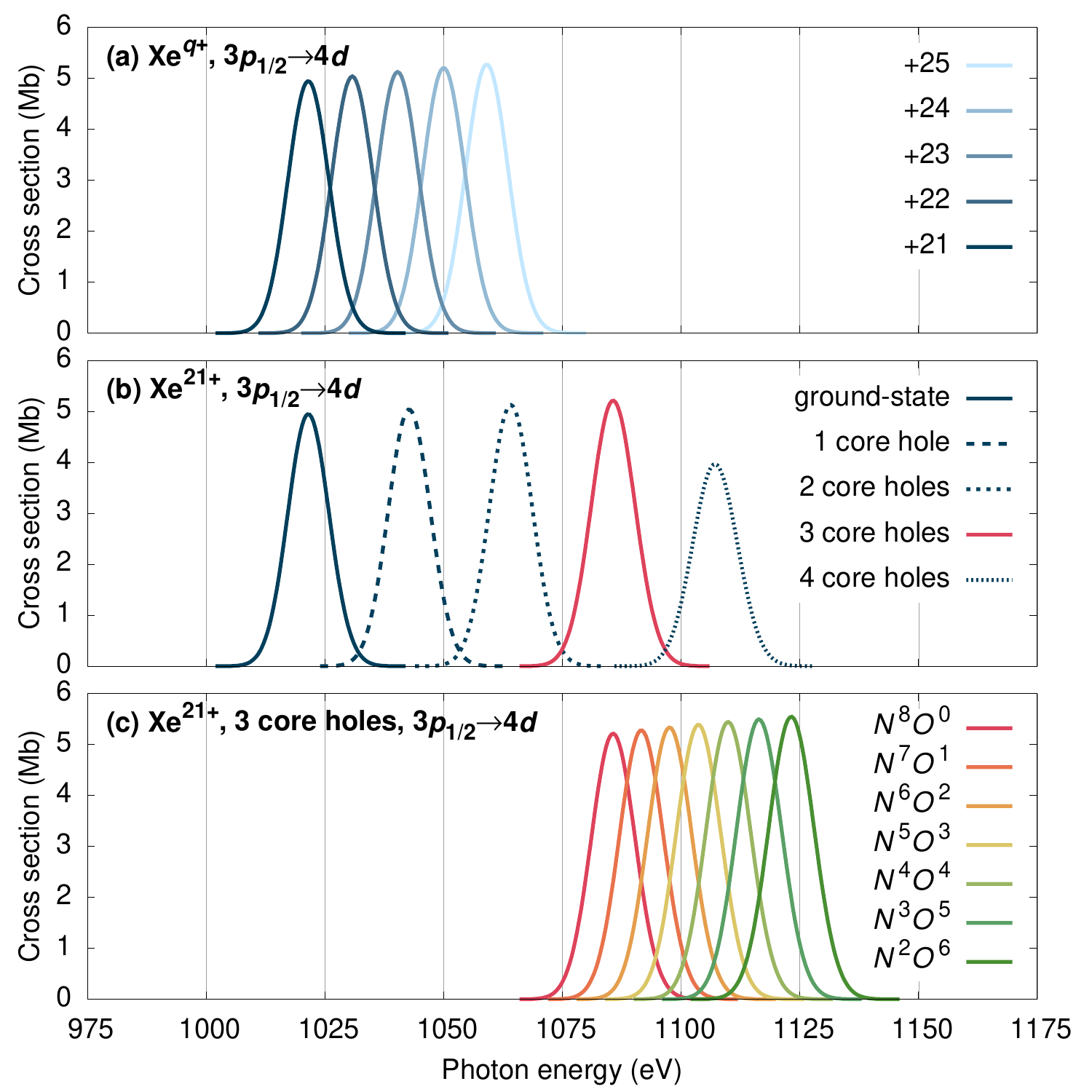}
\caption{Dependence of the calculated photoabsorption cross section of specific \trans{3p_{1/2}}{4d} transitions [peak~D in Fig.~\ref{fig:resonance_profiles}(c)] on (a) charge states (ground electronic configuration), (b) multiple core holes ($3d_{5/2}^{-n}$; $n$ is the number of core holes), and (c) individual valence electronic configurations.
Cross sections were calculated using \textsc{xatom}~\cite{son_impact_2011} and convolved with an energy bandwidth of 1\%.
}
\label{fig:cross_section}
\end{figure}

Supplementary Figure~\ref{fig:cross_section} illustrates how the resonance peaks are sensitive to the electron structure for the case of peak~D in Fig.~\ref{fig:resonance_profiles}(c). The plots are calculated photoabsorption cross sections as a function of photon energy, for a variety of (a) charge states, (b) multiple-core holes, and (c) individual valence electron configurations. The peak of the cross section, corresponding to the \trans{3p_{1/2}}{4d} transition energy, is shifted to lower energy as the charge state decreases. On the other hand, for a fixed charge state, the peak is shifted to higher energy as the number of core holes increases. Lastly, the peak also depends on valence electron configurations ($N^n O^m$ indicates $n$ electrons in the $N$ shell and $m$ electrons in the $O$ shell). Therefore, Supplementary Fig.~\ref{fig:cross_section} demonstrates that ground-state-based assignments can be problematic and a more detailed analysis is crucial.

To obtain a comprehensive picture, we analysed resonant transitions in individual Monte Carlo trajectories corresponding to the selected peaks in Fig.~\ref{fig:resonance_profiles}.
Only the last resonant excitations were analysed because they are most decisive for the final charge state.
Supplementary Table~\ref{table:transition} shows the selected peaks from A to F, specified by the final charge state $q$ and the peak position $E_\text{peak}$.
For each peak, the total number of trajectories used for analysis, $N_\text{tot}$, and the number of trajectories for a specific resonant transition, $N_T$, are listed.
The specific transition $T$ is assigned according to the majority of calculated trajectories, as indicated by $N_T \sim N_\text{tot}$ for all the cases.
The transitions assigned according to the majority are not the same as those from the ground-state-based assignments given in Table~\ref{table:peak_assignment}.

\begin{figure}[H]
\centering
\includegraphics[width=\textwidth]{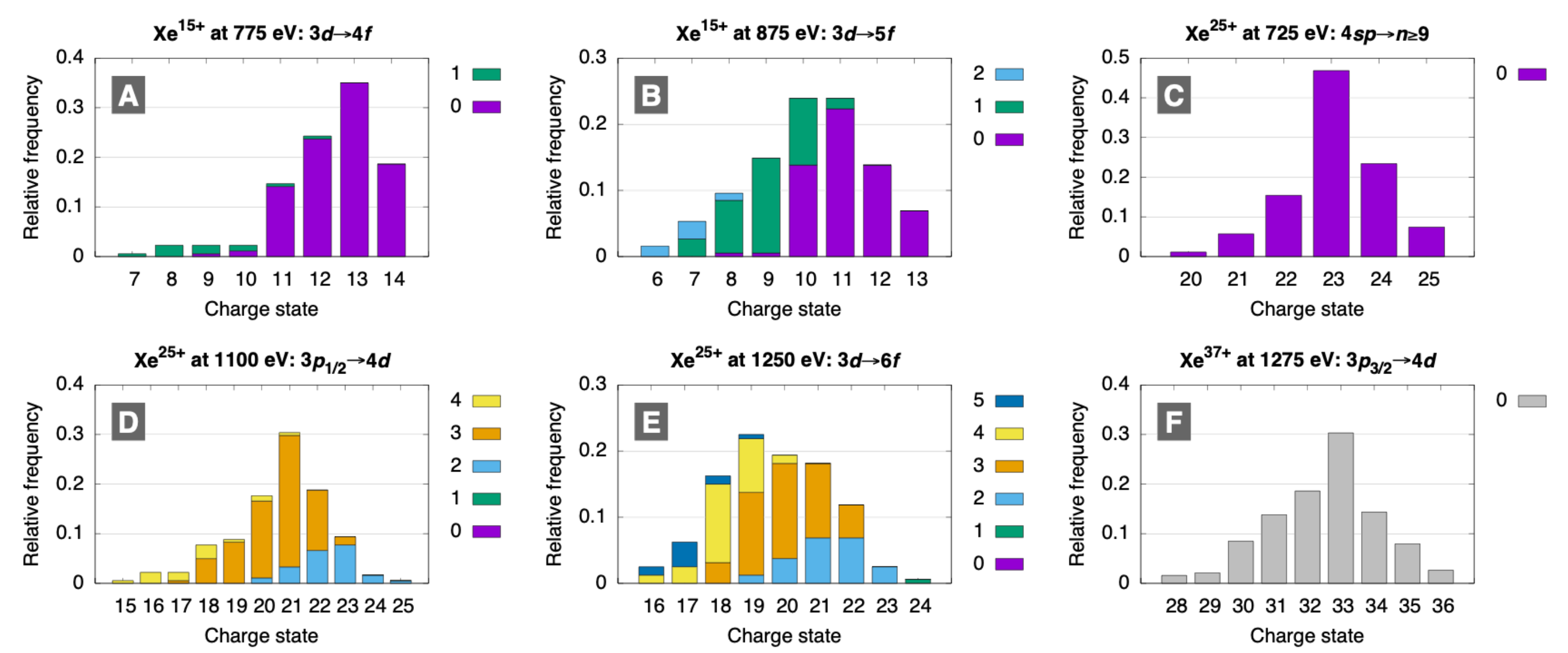}
\caption{Analysis of specific resonant excitations for different final charge states and photon energies, corresponding to peaks A--F in Fig.~\ref{fig:resonance_profiles}. The dominant last resonant excitation is specified in the title of each panel. Different colours indicate the relative number of $M$-shell core holes present at the time of the resonant excitations.}
\label{fig:histogram_last_resonances_all}
\end{figure}

The electronic structure when the last resonant excitation takes place is reflected by the charge state and the number of $M$-shell core holes at the time of the respective transition.
Supplementary Figure~\ref{fig:histogram_last_resonances_all} shows normalised histograms of the last resonant excitations for peaks A--F, analysed by the precursor charge state and the number of $M$-shell core holes, in the same way as was done in Fig.~\ref{fig:histogram_last_resonances}.
Panel~E is identical to Fig.~\ref{fig:histogram_last_resonances}(a).
Panels~A and B confirm that peaks~A and B in Fig.~\ref{fig:resonance_profiles}(a) originate from the \trans{3d}{4f} and \trans{3d}{5f} transitions, respectively, and demonstrate that single and double core holes are present at lower charge states when the respective resonant excitation happens.
As indicated in the title of panel~C, peak~C in Fig.~\ref{fig:resonance_profiles}(c) corresponds to a transition from the $N$ shell ($n$=4), the outermost shell for $+8\!\le\!q\!<\!+26$. 
In the case of panels~D and E, our trajectory analysis reveals that the \trans{3p_{1/2}}{4d} and \trans{3d}{6f} transitions are responsible for peaks~D and E in Fig.~\ref{fig:resonance_profiles}(c). They are created exclusively via multiple core holes, which explains why the ground-state-based assignment fails. 
For peak~F in Fig.~\ref{fig:resonance_profiles}(d), the $M$ shell is the outermost shell for $q$$\ge$$+26$; therefore, no $M$-shell \textit{core} holes exist, as indicated by the grey colour in panel~F.

\newpage
\section*{Supplementary Figures}

\begin{figure}[H]
\centering
\includegraphics[width=0.70\figurewidth]{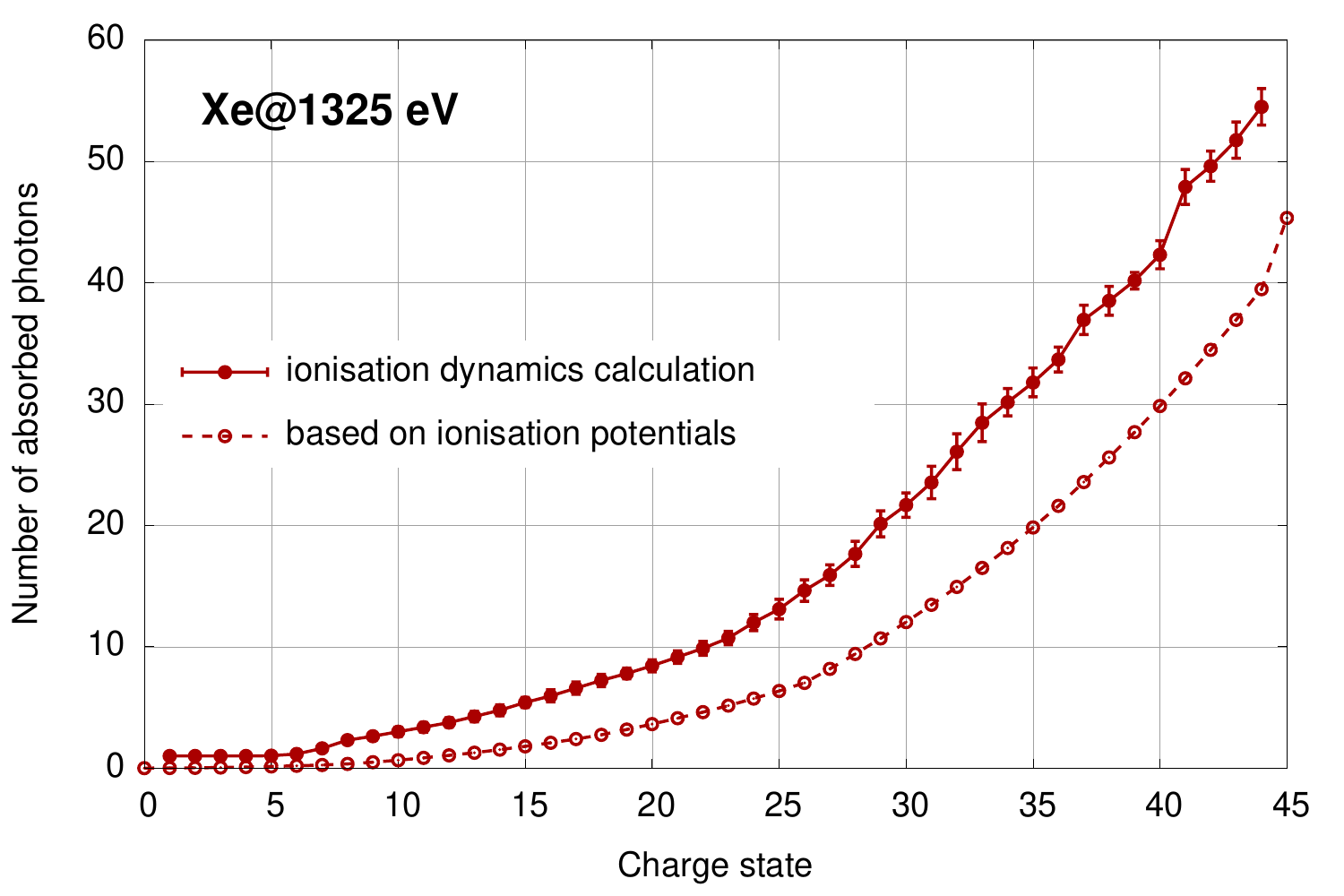}
\caption{Number of X-ray photons required to reach a given charge state of Xe at a photon energy of 1325\,eV. The dashed line was calculated as the sum of ionisation potentials divided by the photon energy of 1325\,eV, which indicates the minimum amount of energy (number of photons) needed to create the respective charge state. The solid line indicates the mean value of the number of photons that are actually absorbed during X-ray multiphoton ionisation dynamics, as calculated using \textsc{xatom}, where the upper and lower bounds represent the standard deviations of the distributions of the number of absorbed photons. The number of absorbed photons in our ionisation model significantly exceeds the minimum number of photons based on the ionisation potentials.
}
\label{fig:N_abs_ph}
\end{figure}

\begin{figure}[H]
\begin{center}
\includegraphics[width=0.70\figurewidth]{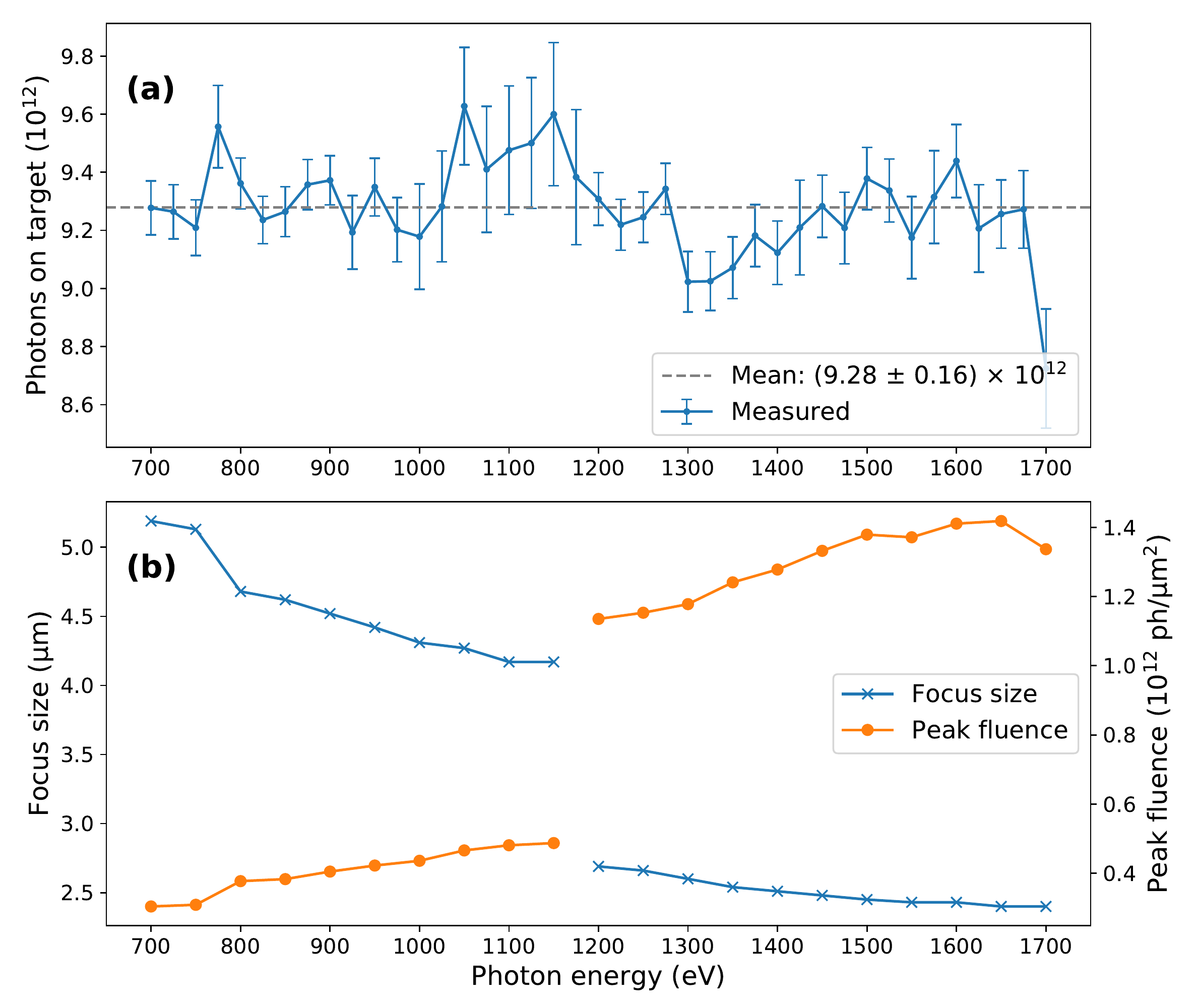}
\caption{(a) Number of photons on target, as recorded by the gas-monitor detector downstream of the experiment as a function of photon energy. The error bars represent the statistical uncertainties. (b) Peak fluence (orange, right axis) and focus size (blue, left axis) were obtained from the fluence calibration with argon (see text in Supplementary Discussion~\ref{app:volume}).}
\label{fig:num_photons}
\end{center}
\end{figure}

\newpage
\section*{Supplementary Tables}

\begin{table}[h]
\caption{\label{table:E_shift}%
Transition energies for exemplary charge states (ground electronic configuration), comparing calculations using the \textsc{fac}~\cite{gu_flexible_2008} and \textsc{xatom} toolkits~\cite{son_impact_2011,jurek_xmdyn_2016}. $\Delta = E_\text{FAC} - E_\text{XATOM}$.
All level-resolved \textsc{fac} results are averaged in order to be compared with configuration-resolved \textsc{xatom} results. The state-specific transition $3d^{-1}_{3/2} 4f^1_{5/2} (J^P=1^-)$ at \Xe{26}, which is the most probable transition for that charge state, can be compared with the available experimental data~\cite{kramida_nist_1999} (EXP).
While \textsc{fac} provides a reliable transition energy in comparison with the experimental result, the comparison between \textsc{xatom} and \textsc{fac} suggests that there is a systematic shift in \textsc{xatom} for \trans{3d}{nf}, regardless of the charge state: $\Delta$=$-22.6\pm1.7$\,eV.
However, the shifts for the $3p_{1/2}$ and $3p_{3/2}$ transitions are different. 
To obtain better accuracy, one needs an improved electronic structure theory, such as many-body perturbation theory~\cite{budewig_theoretical_2022}, beyond the mean-field approach that is employed in this work.
}
\centering
\begin{tabular}{rlrrrr}
\toprule
$q$	& Transition 						& EXP		& \textsc{fac}		& \textsc{xatom}		& $\Delta$	\\
\midrule
$+8$	& $3d_{5/2} \rightarrow 5f_{5/2}$	&			& 736.52	& 757.80	& $-21.28$	\\
		& $3d_{5/2} \rightarrow 5f_{7/2}$	&			& 736.72	& 757.88	& $-21.16$	\\
		& $3d_{3/2} \rightarrow 5f_{5/2}$	&			& 749.47	& 771.33	& $-21.86$	\\
\midrule
$+18$	& $3d_{5/2} \rightarrow 4f_{5/2}$	&			& 772.44	& 794.44	& $-22.00$	\\
		& $3d_{5/2} \rightarrow 4f_{7/2}$	&			& 773.46	& 795.07	& $-21.61$	\\
		& $3d_{3/2} \rightarrow 4f_{5/2}$	&			& 787.07	& 808.37	& $-21.30$	\\
\vspace{-3mm} \\
		& $3p_{3/2} \rightarrow 4d_{3/2}$	&			& 932.45	& 937.53	& $-5.08$	\\
		& $3p_{3/2} \rightarrow 4d_{5/2}$	&			& 935.96	& 940.83	& $-4.87$	\\
		& $3p_{1/2} \rightarrow 4d_{3/2}$	&			& 996.49	& 995.60	& $0.89$	\\
\vspace{-3mm} \\
		& $3d_{5/2} \rightarrow 5f_{5/2}$	&			& 916.85	& 941.16	& $-24.31$	\\
		& $3d_{5/2} \rightarrow 5f_{7/2}$	&			& 917.35	& 941.43	& $-24.08$	\\
		& $3d_{3/2} \rightarrow 5f_{5/2}$	&			& 930.47	& 955.09	& $-24.62$	\\
\midrule
$+26$	& $3d_{5/2} \rightarrow 4f_{5/2}$	&			& 841.21	& 864.18	& $-22.97$	\\
		& $3d_{5/2} \rightarrow 4f_{7/2}$	&			& 842.55	& 865.14	& $-22.59$	\\
		& $3d_{3/2} \rightarrow 4f_{5/2}$	&			& 857.29	& 878.66	& $-21.37$	\\
        & $3d^{-1}_{3/2} 4f^1_{5/2} (1^-)$  & 870.25	& 870.24	& 			& 			\\
		\vspace{-3mm} \\
		& $3p_{3/2} \rightarrow 4d_{3/2}$	&			& 1006.09	& 1006.67	& $-0.58$	\\
		& $3p_{3/2} \rightarrow 4d_{5/2}$	&			& 1010.43	& 1010.88	& $-0.45$	\\
		& $3p_{1/2} \rightarrow 4d_{3/2}$	&			& 1071.77	& 1066.47	& $5.30$	\\
\vspace{-3mm} \\
		& $3d_{5/2} \rightarrow 5f_{5/2}$	&			& 1077.77	& 1098.98	& $-21.12$	\\
		& $3d_{5/2} \rightarrow 5f_{7/2}$	&			& 1078.56	& 1099.50	& $-20.94$	\\
		& $3d_{3/2} \rightarrow 5f_{5/2}$	&			& 1092.26	& 1113.46	& $-21.20$	\\
\bottomrule
\end{tabular}
\end{table}

\begin{table}
\caption{\label{table:peak_assignment}%
Peak assignment based on the ground-state transition energies of charge state $q\!-\!1$.
All energies are in eV.
Parentheses are used if the energy difference is $\ge$40\,eV.
Note that there is no transition from $3d$ for $q\!>\!+36$.
}
\centering
\begin{tabular}{rrccccccc}
\toprule
$q$     & $E_\text{peak}$ & Label   & \trans{3d}{4f}    & \trans{3d}{5f}    & \trans{3d}{6f}    & \trans{3d}{7f}    & \trans{3p_{3/2}}{4d}    & \trans{3p_{1/2}}{4d} \\
\midrule
$+15$   & 775   & A  & 763   & --    & --    & --    & --    & --    \\
        & 875   & B  & --    & 866   & 914   & (942) & 912   &(967)  \\
\midrule
$+21$   & 825   & -- & 813   & --    & --    & --    & --    & --    \\
        & 975   & -- & --    & 981   & --    & --    & 958   & --    \\
        & 1125  & -- & --    & --    &(1067) & 1117  & --    &(1013) \\
\midrule
$+25$   & 725   & C  & --    & --    & --    & --    & --    & --    \\
        & 875   & -- & 849   & --    & --    & --    & --    & --    \\
        & 1025  & -- & --    & --    & --    & --    & 995   & 1050  \\
        & 1100  & D  & --    &(1060) & --    & --    & --    & --    \\
        & 1250  & E  & --    & --    &(1174) & 1241  & --    & --    \\
\midrule
$+37$   & 1175  & -- & --    & --    & --    & --    & --    & --    \\
        & 1275  & F  & --    & --    & --    & --    & 1308  &(1375) \\
\bottomrule
\end{tabular}
\end{table}

\begin{table}
\caption{\label{table:transition}
Resonant transitions that are analysed in Fig.~\ref{fig:histogram_last_resonances_all}, which correspond to the majority of calculated Monte Carlo trajectories for the final charge state $q$ and photon energy $E_\text{peak}$ (in eV) in the resonance spectrum of Fig.~\ref{fig:resonance_profiles}.
$N_\text{tot}$ is the total number of trajectories that give rise to the final charge state $q$ at $E_\text{peak}$, and $N_T$ is the number of trajectories for the specific transition $T$ that dominantly occurs at the last resonant excitation among the $N_\text{tot}$ trajectories.
Where possible, the transitions as a result of the ground-state-based peak assignment given in Table~\ref{table:peak_assignment} are included.
}
\centering
\begin{tabular}{@{\extracolsep{\fill}}crrrrll@{\extracolsep{\fill}}}
\toprule
  &  &  & \multicolumn{3}{@{}c@{}}{Monte Carlo analysis} &  \\\cmidrule{4-6}
Label & $q$	& $E_\text{peak}$ & $N_\text{tot}$ & $N_T$ & Transition $T$	& Assigned in Table~\ref{table:peak_assignment}  \\
\midrule
A & +15	& 775		& 180 & 177	& \trans{3d}{4f} 		  & \qquad \trans{3d}{4f}   \\
B & +15	& 875		& 209 & 188	& \trans{3d}{5f} 		  & \qquad \trans{3d}{5f}   \\
C & +25	& 725		& 194 & 175	& \trans{4sp}{n\!\geq\!9} & \qquad --               \\
D & +25	& 1100		& 196 & 181	& \trans{3p_{1/2}}{4d}    & \qquad --               \\
E & +25	& 1250		& 201 & 160	& \trans{3d}{6f}          & \qquad \trans{3d}{7f}   \\
F & +37	& 1275		& 270 & 188	& \trans{3p_{3/2}}{4d}    & \qquad \trans{3p_{3/2}}{4d}   \\
\bottomrule
\end{tabular}
\end{table}

\newcommand{\config}[4]{[Ne]#1#2#3#4}
\newcommand{\shell}[6]{\ensuremath{%
\ifthenelse{\equal{#2}{0}}{}{#1s^{#2}}%
\ifthenelse{\equal{#3}{0}}{}{#1p_{1/2}^{#3}}%
\ifthenelse{\equal{#4}{0}}{}{#1p_{3/2}^{#4}}%
\ifthenelse{\equal{#5}{0}}{}{#1d_{3/2}^{#5}}%
\ifthenelse{\equal{#6}{0}}{}{#1d_{5/2}^{#6}}}}
\newcommand{\Mshell}[5]{\ \ensuremath{3s[{#1}]3p[{#2},{#3}]3d[{#4},{#5}]}\ }
\newcommand{\Nshell}[5]{\shell{4}{#1}{#2}{#3}{#4}{#5}}
\newcommand{\Oshell}[3]{\shell{5}{#1}{#2}{#3}{0}{0}}
\begin{table}
\caption{\label{table:configuration}
Calculated lifetimes of core-hole states of Xe that are formed in the course of the exemplary Monte Carlo trajectory in Fig.~\ref{fig:histogram_last_resonances}(b).
For the electron configuration, [Ne] means $1s^2 2s^2 2p_{1/2}^2 2p_{3/2}^4$ and $3s[n_1] 3p[n_2,n_3] 3d[n_4,n_5]$ refers to $3s^{n_1} 3p_{1/2}^{n_2} 3p_{3/2}^{n_3} 3d_{3/2}^{n_4} 3d_{5/2}^{n_5}$. $N_M$ indicates the number of holes in the $M$ shell.
}
\centering
\begin{tabular}{rlrr}
\toprule
$q$   & Electron configuration & $N_M$ 		& Lifetime (fs) \\
\midrule
$+0$  & [Ne]\Mshell{2}{2}{4}{4}{6}\Nshell{2}{2}{4}{4}{6}\Oshell{2}{2}{4} & 0 & --\phantom{.000} \\
$+1$  & [Ne]\Mshell{1}{2}{4}{4}{6}\Nshell{2}{2}{4}{4}{6}\Oshell{2}{2}{4} & 1 & 0.054 \\
$+2$  & [Ne]\Mshell{2}{2}{3}{4}{6}\Nshell{2}{2}{4}{4}{6}\Oshell{2}{2}{3} & 1 & 0.13\phantom{0} \\
$+3$  & [Ne]\Mshell{2}{2}{4}{3}{6}\Nshell{2}{2}{4}{4}{5}\Oshell{2}{2}{3} & 1 & 1.0\phantom{00} \\
$+4$  & [Ne]\Mshell{2}{2}{4}{4}{6}\Nshell{2}{2}{4}{2}{5}\Oshell{2}{2}{3} & 0 & 5.1\phantom{00} \\
$+5$  & [Ne]\Mshell{2}{2}{4}{4}{5}\Nshell{2}{2}{4}{2}{5}\Oshell{2}{2}{3} & 1 & 1.1\phantom{00} \\
$+6$  & [Ne]\Mshell{2}{2}{4}{4}{5}\Nshell{1}{2}{4}{2}{5}\Oshell{2}{2}{3} & 1 & 0.42\phantom{0} \\
$+7$  & [Ne]\Mshell{2}{2}{4}{4}{4}\Nshell{1}{2}{4}{2}{5}\Oshell{2}{2}{3} & 2 & 0.31\phantom{0} \\
$+8$  & [Ne]\Mshell{2}{2}{4}{4}{4}\Nshell{2}{2}{4}{2}{4}\Oshell{2}{2}{2} & 2 & 0.78\phantom{0} \\
$+9$  & [Ne]\Mshell{2}{2}{4}{4}{3}\Nshell{2}{2}{4}{2}{4}\Oshell{2}{2}{2} & 3 & 0.50\phantom{0} \\
$+10$ & [Ne]\Mshell{2}{2}{4}{4}{4}\Nshell{2}{2}{4}{2}{2}\Oshell{2}{2}{2} & 2 & 1.5\phantom{00} \\
$+11$ & [Ne]\Mshell{2}{2}{4}{4}{5}\Nshell{2}{2}{4}{1}{1}\Oshell{2}{2}{2} & 1 & 5.7\phantom{00} \\
$+12$ & [Ne]\Mshell{2}{2}{4}{3}{5}\Nshell{2}{2}{4}{1}{1}\Oshell{2}{2}{2} & 2 & 2.4\phantom{00} \\
$+13$ & [Ne]\Mshell{2}{2}{4}{2}{5}\Nshell{2}{2}{4}{1}{1}\Oshell{2}{2}{2} & 3 & 1.5\phantom{00} \\
$+14$ & [Ne]\Mshell{2}{2}{4}{1}{5}\Nshell{2}{2}{4}{1}{1}\Oshell{2}{2}{2} & 4 & 1.1\phantom{00} \\
$+14$ & [Ne]\Mshell{2}{2}{3}{1}{5}\Nshell{2}{2}{4}{1}{1}\Oshell{2}{2}{2}$5d_{3/2}^1$ & 5 & 0.54\phantom{0} \\
$+15$ & [Ne]\Mshell{2}{2}{3}{2}{5}\Nshell{2}{1}{4}{0}{1}\Oshell{2}{2}{2}$5d_{3/2}^1$ & 4 & 0.82\phantom{0} \\
$+16$ & [Ne]\Mshell{2}{2}{3}{2}{6}\Nshell{2}{0}{3}{0}{1}\Oshell{2}{2}{2}$5d_{3/2}^1$ & 3 & 1.0\phantom{00} \\
$+16$ & [Ne]\Mshell{2}{2}{3}{2}{5}\Nshell{2}{0}{3}{0}{1}\Oshell{2}{2}{2}$5d_{3/2}^1 11f_{7/2}^1$ & 4 & 0.88\phantom{0} \\
$+16$ & [Ne]\Mshell{2}{2}{3}{2}{4}\Nshell{2}{0}{3}{0}{1}\Oshell{2}{2}{2}$5d_{3/2}^1 7f_{7/2}^1 11f_{7/2}^1$ & 5 & 0.67\phantom{0} \\
$+17$ & [Ne]\Mshell{2}{2}{3}{2}{4}\Nshell{2}{1}{3}{0}{1}\Oshell{2}{1}{2}$5d_{3/2}^1 11f_{7/2}^1$ & 5 & 1.1\phantom{00} \\
$+17$ & [Ne]\Mshell{2}{2}{3}{2}{3}\Nshell{2}{1}{3}{0}{1}\Oshell{2}{1}{2}$5d_{3/2}^1 6f_{7/2}^1 11f_{7/2}^1$ & 6 & 0.75\phantom{0} \\
$+18$ & [Ne]\Mshell{2}{2}{3}{2}{3}\Nshell{2}{1}{4}{0}{1}\Oshell{2}{1}{1}$6f_{7/2}^1 11f_{7/2}^1$ & 6 & 0.80\phantom{0} \\
$+19$ & [Ne]\Mshell{2}{2}{3}{2}{4}\Nshell{1}{1}{4}{0}{0}\Oshell{2}{1}{1}$6f_{7/2}^1 11f_{7/2}^1$ & 5 & 1.3\phantom{00} \\
$+20$ & [Ne]\Mshell{2}{2}{4}{2}{4}\Nshell{1}{1}{2}{0}{0}\Oshell{2}{1}{1}$6f_{7/2}^1 11f_{7/2}^1$ & 4 & 5.5\phantom{00} \\
$+21$ & [Ne]\Mshell{2}{2}{4}{2}{4}\Nshell{1}{1}{2}{0}{0}\Oshell{1}{1}{1}$6f_{7/2}^1 11f_{7/2}^1$ & 4 & 7.4\phantom{00} \\
$+22$ & [Ne]\Mshell{2}{2}{4}{2}{4}\Nshell{1}{0}{2}{0}{0}\Oshell{1}{1}{1}$6f_{7/2}^1 11f_{7/2}^1$ & 4 & 6.9\phantom{00} \\
$+23$ & [Ne]\Mshell{2}{2}{4}{2}{4}\Nshell{1}{1}{2}{0}{0}\Oshell{1}{0}{1}$11f_{7/2}^1$ & 4 & 32\phantom{.000} \\
$+24$ & [Ne]\Mshell{2}{2}{4}{2}{5}\Nshell{1}{0}{2}{0}{0}\Oshell{1}{0}{0}$11f_{7/2}^1$ & 3 & 130\phantom{.000} \\
$+24$ & [Ne]\Mshell{2}{2}{4}{2}{5}\Nshell{1}{1}{2}{0}{0}\Oshell{0}{0}{0}$11f_{7/2}^1$ & 3 & 620\phantom{.000} \\
$+25$ & [Ne]\Mshell{2}{2}{4}{2}{6}\Nshell{1}{1}{1}{0}{0}\Oshell{0}{0}{0} & 2 & 1600\phantom{.000} \\
$+25$ & [Ne]\Mshell{2}{2}{4}{3}{6}\Nshell{1}{0}{1}{0}{0}\Oshell{0}{0}{0} & 1 & 20000\phantom{.000} \\
$+25$ & [Ne]\Mshell{2}{2}{4}{4}{6}\Nshell{1}{0}{0}{0}{0}\Oshell{0}{0}{0} & 0 & --\phantom{.000} \\
\bottomrule
\end{tabular}
\end{table}

\end{appendix}

\end{document}